\documentclass[
10pt,
a4paper,
superscriptaddress,
reprint,
twocolumn,
amsmath,
amssymb,
amsfonts,
aip,
cha,
showkeys,
floatfix,
]{revtex4-2}

\usepackage{graphicx}%
\usepackage{dcolumn}%
\usepackage{bm}%

\usepackage[separate-uncertainty=true]{siunitx}
\usepackage{physics}
\AtBeginDocument{\RenewCommandCopy\qty\SI}
\DeclareSIUnit\sq{\ensuremath\Box}
\DeclareSIUnit{\sqrthz}{\ensuremath{\sqrt{\mathrm{\hertz}}}}
\DeclareSIUnit{\sample}{S}

\usepackage{hyperref}
\usepackage{xcolor}
\hypersetup{
    colorlinks,
    linkcolor={red!50!black},
    citecolor={blue!50!black},
    urlcolor={blue!80!black}
}

\newcommand{\meff}{\ensuremath{ m_{\mathrm{eff}} }}
\newcommand{\Ceff}{\ensuremath{ \mathcal{C}_{\mathrm{eff}} }}

\newcommand{\kB}{\ensuremath{ k_{\mathrm{B}} } }

\newcommand{\Kerr}{\ensuremath{ \mathcal{K} } }

\newcommand{\Fmin}{\ensuremath{ F_{\mathrm{min}} }}

\newcommand{\ncav}{\ensuremath{ \bar{n}_{\mathrm{c}} } }
\newcommand{\nadd}{\ensuremath{ n_{\mathrm{add}} } }
\newcommand{\nmech}{\ensuremath{ \bar{n}_{\mathrm{m}} } }

\begin{document}

\preprint{arXiv:0000.00000}

\title{Intrinsic Kerr amplification for microwave electromechanics}

\author{Ermes Scarano}
\affiliation{Department of Applied Physics, KTH Royal Institute of Technology, Hannes Alfvéns väg 12, SE-114 19 Stockholm, Sweden}
\author{Elisabet K. Arvidsson}
\affiliation{Department of Applied Physics, KTH Royal Institute of Technology, Hannes Alfvéns väg 12, SE-114 19 Stockholm, Sweden}
\author{August K. Roos}
\affiliation{Department of Applied Physics, KTH Royal Institute of Technology, Hannes Alfvéns väg 12, SE-114 19 Stockholm, Sweden}
\author{Erik Holmgren}
\affiliation{Department of Applied Physics, KTH Royal Institute of Technology, Hannes Alfvéns väg 12, SE-114 19 Stockholm, Sweden}
\author{David B. Haviland}
\email{haviland@kth.se}
\affiliation{Department of Applied Physics, KTH Royal Institute of Technology, Hannes Alfvéns väg 12, SE-114 19 Stockholm, Sweden}
\date{June 14, 2024}%

\begin{abstract}

Electromechanical transduction gain of \SI{21}{\decibel} is realized in a micro-cantilever resonant force sensor operated in the unresolved-sideband regime. Strain-dependent kinetic inductance weakly couples cantilever motion to a superconducting nonlinear resonant circuit. A single pump generates motional sidebands and parametrically amplifies them via four-wave mixing. We study the gain and added noise, and we analyze potential benefits of this integrated amplification process in the context force sensitivity.

\end{abstract}

\keywords{kinetic inductance; parametric amplification; four-wave mixing; optomechanics; force sensing; strain}

\maketitle

A resonant electromechanical transducer converts force to motion with a responsivity solely determined by its mechanical design. Cavity optomechanics offers an efficient scheme for measuring motion, but in many cases the transducer's overall sensitivity to force is limited by added noise in the measurement chain. In such cases the addition of a quantum-limited parametric amplifier can enhance sensitivity~\cite{teufel2009nanomechanicalmotion}. Here we use the intrinsic Kerr nonlinearity of a superconducting microwave resonant circuit to achieve transduction gain in a micro-cantilever force sensor.  However, for the device studied here, analysis of the noise revealed that this gain did not result in enhanced sensitivity to force. 

In comparison to optical cavities, microwave circuits are appealing for their relative ease of integration with micro- and nano-electromechanical systems and their ability to achieve stronger electromechanical coupling. Since their first implementation~\cite{regal2008nanomechanicalmotion, teufel2009nanomechanicalmotion}, a large variety of superconducting microwave electromechanical devices have been studied and optimized for different goals~\cite{aspelmeyer2014review}, such as a large single-photon coupling rate $g_{0}$~\cite{nation2016ultrastrongcoupling, reed2017strongcoupling, rodrigues2019couplingmicrowavephotons, zoepfl2020singlephotoncooling}, electromechanical cooperativity $\mathcal{C}$~\cite{peterson2019strongcoupling, ren2020highquantumcooperativity, bozkurt2023quantum}, amplification~\cite{cattiaux2020beyondlinearcoupling, shin2022frequencycombs}, and efficient cooling of the mechanical mode~\cite{blencowe2007squidanalysis, teufel2011sidebandcooling, safavi-naeini2012nanomechanicalresonator, yuan2015cavity3d, cattiaux2021groundstatecooling, bothner2022kerroptomechanics, seis2022cooling}. Depending on the application, navigation of the design parameter space results in significantly different paths toward the goal.

Optimizing an electromechanical transducer for sensitivity to force constrains the mechanical resonator's shape, size, mass and resonance frequency. These constraints can conflict with other figures of merit. For example, a larger mass reduces the coupling rate $g_{0}$ and mechanical resonance frequency $\Omega_{m}$, leading to operation in the unresolved-sideband regime and larger phonon occupation $\nmech$ at a given temperature.  Reduced $g_{0}$ can be compensated for with a stronger pump that increases the intracavity photon number $\ncav$, but superconducting circuits experience current-induced depairing that sets a limit on $\ncav$.  This depairing results in a nonlinear inductance which is usually seen as a problem~\cite{yuan2015cavity3d}. However, recent works exploit this nonlinearity for efficient cooling of the mechanical mode, both in the resolved~\cite{bothner2022kerroptomechanics} and unresolved~\cite{zoepfl2023kerrenhancedcooling} sideband regimes. The Kerr nonlinearity of kinetic inductance is also used for parametric amplification~\cite{zmuidzinas2012review} via three-wave or four-wave mixing processes~\cite{vissers2016twpa, erickson2017theorykinetictwpa}. 

In this paper we report on a microwave resonant circuit patterned from a thin film of niobium titanium nitride (Nb-Ti-N), which is weakly coupled to a micro-cantilever in the unresolved-sideband regime. We exploit the intrinsic Kerr nonlinearity of the superconducting circuit in a four-wave mixing process to realize \SI{21}{\decibel} of transduction gain.  We define the transduction gain as the ratio of the measured amplitude of the optomechanical motional sideband, normalized to that which would be produced by a linear cavity pumped to the same intracavity photon number.  

In the cavity optomechanical transduction scheme, mechanical fluctuations with average phonon number $\nmech$ are imprinted on the optical spectrum leaking out of the cavity, appearing as motional noise sidebands around the cavity pump frequency. In microwave electromechanics, the signal leaking from a resonant circuit into a transmission line is amplified before being demodulated and typically the added noise of this amplifier is the limiting factor degrading force sensitivity. Force sensitivity is expressed as a minimum detectable force $\Fmin$, i.e. the force per unit bandwidth detected at unity signal-to-noise ratio. This definition accounts for actual force noise from the environment and the backaction of measurement, as well as added photon shot noise and amplification noise $\nadd$ expressed as an equivalent force noise~\cite{bowen2015quantumoptomechanics}. The normalized force sensitivity is given by
\small
\begin{equation}
    \frac{\Fmin}{F_{\mathrm{SQL}}} = \sqrt{\underbrace{\left(\nmech+\frac{1}{2}\right)}_{\text{Mechanics}} + \underbrace{\lvert \Ceff \rvert}_{\text{Backaction}} +\underbrace{\frac{1}{16 \lvert \Ceff \rvert}}_{\text{Shot noise}}+\underbrace{\frac{\nadd}{8 \lvert \Ceff \rvert}}_{\text{Amplifier noise}}},
    \label{eqn:minimum_detectable_force}
\end{equation}
\normalsize
where $F_{\mathrm{SQL}} = \sqrt{2\meff\Gamma\hbar\Omega_{m}}$ is the force sensitivity at the standard quantum limit (SQL), set by the resonators internal loss rate $\Gamma$ and effective mass $\meff$. Measurement induces noise, which translates to force through the effective cooperativity $\Ceff$~\cite{bowen2015quantumoptomechanics}. In the unresolved-sideband regime the effective cooperativity can be approximated to,
\begin{equation}
    \Ceff(\ncav)= \frac{1}{(1-2i\omega/\kappa)^2}\frac{4 g_{0}^{2} \ncav}{\kappa \Gamma} \simeq \frac{4 g_{0}^{2} \ncav}{\kappa \Gamma},
    \label{eqn:effective_cooperativity}
\end{equation}
where $\kappa$ is total loss rate of the cavity. 
\begin{figure}
    \centering
    \includegraphics[width=\linewidth]{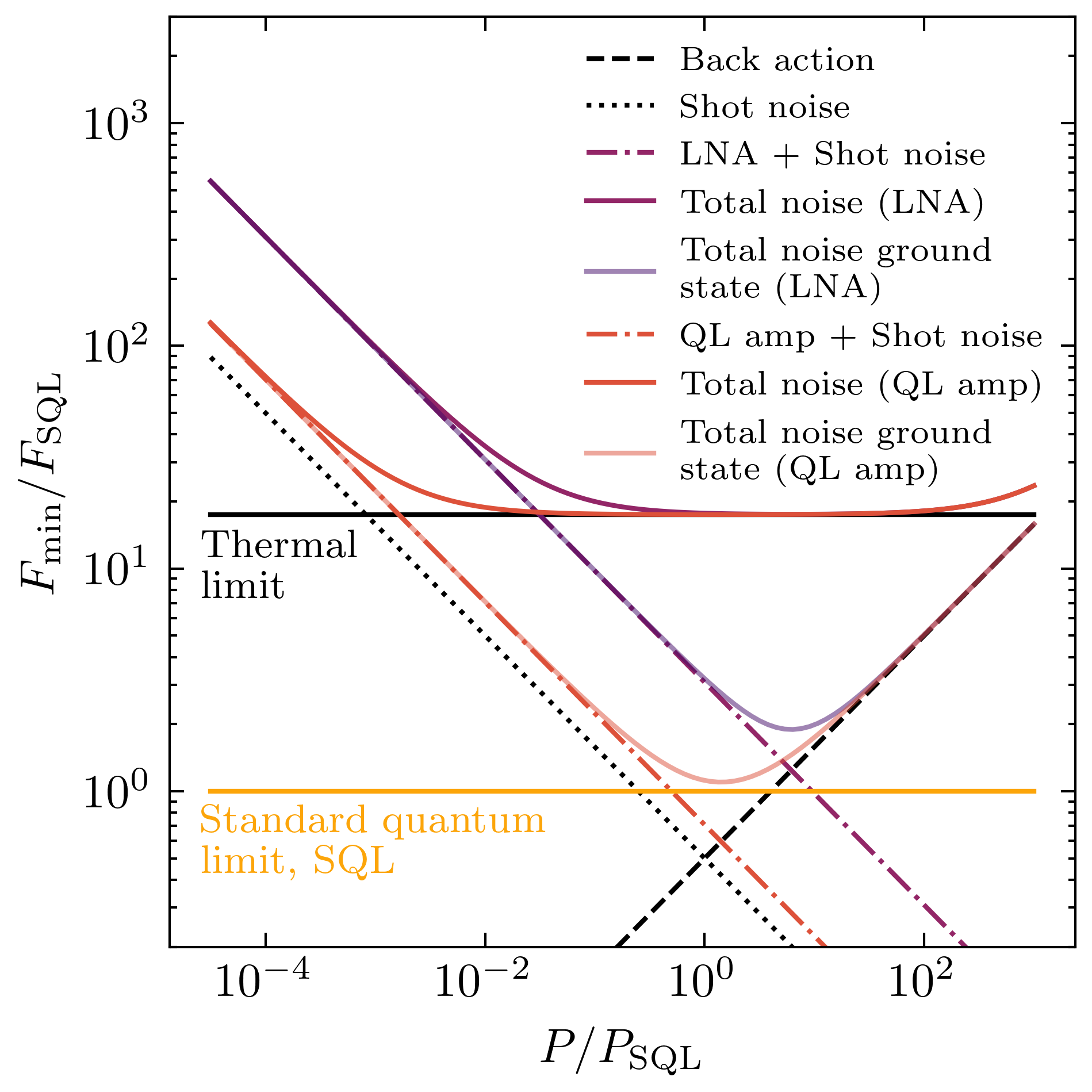}
    \caption{\label{fig:standard_quantum_limit}
        Force sensitivity on mechanical resonance versus input power, both normalized to that at the standard quantum limit. The total force sensitivity (solid lines) is given by the mechanical fluctuations, backaction noise, as well as the force-equivalent shot noise and added amplifier noise. The effect of an additional stage of amplification is shown, either for a thermal state ($\bar{n}_m = 304$) or the quantum ground state of the mechanical mode.
    }
\end{figure} 
Figure~\ref{fig:standard_quantum_limit} shows each contribution of Eqn.~\eqref{eqn:minimum_detectable_force} as a function of input power $P$ and their overall effect on sensitivity. Here $P_{\mathrm{SQL}}$ is the applied microwave power when the measurement is at the standard quantum limit, where shot noise and backaction noise are equal, corresponding to $\Ceff$ = 1/4. For input power below $P_{\mathrm{SQL}}$, the added noise of the first stage of amplification limits force sensitivity. For example, in Fig.~\ref{fig:standard_quantum_limit} we show the contribution from a cryogenic low-noise amplifier (LNA) with an equivalent noise temperature of \SI{4}{\kelvin}. At \SI{4.5}{\giga\hertz} this correspond to approximately $n_\mathrm{LNA}=19$~photons, deteriorating the force sensitivity so that detection at SQL is no longer possible and the optimal sensitivity is achieved for powers $P > P_{\mathrm{SQL}}$.

\begin{figure}[ht]
    \centering
    \includegraphics[width=0.95\linewidth]{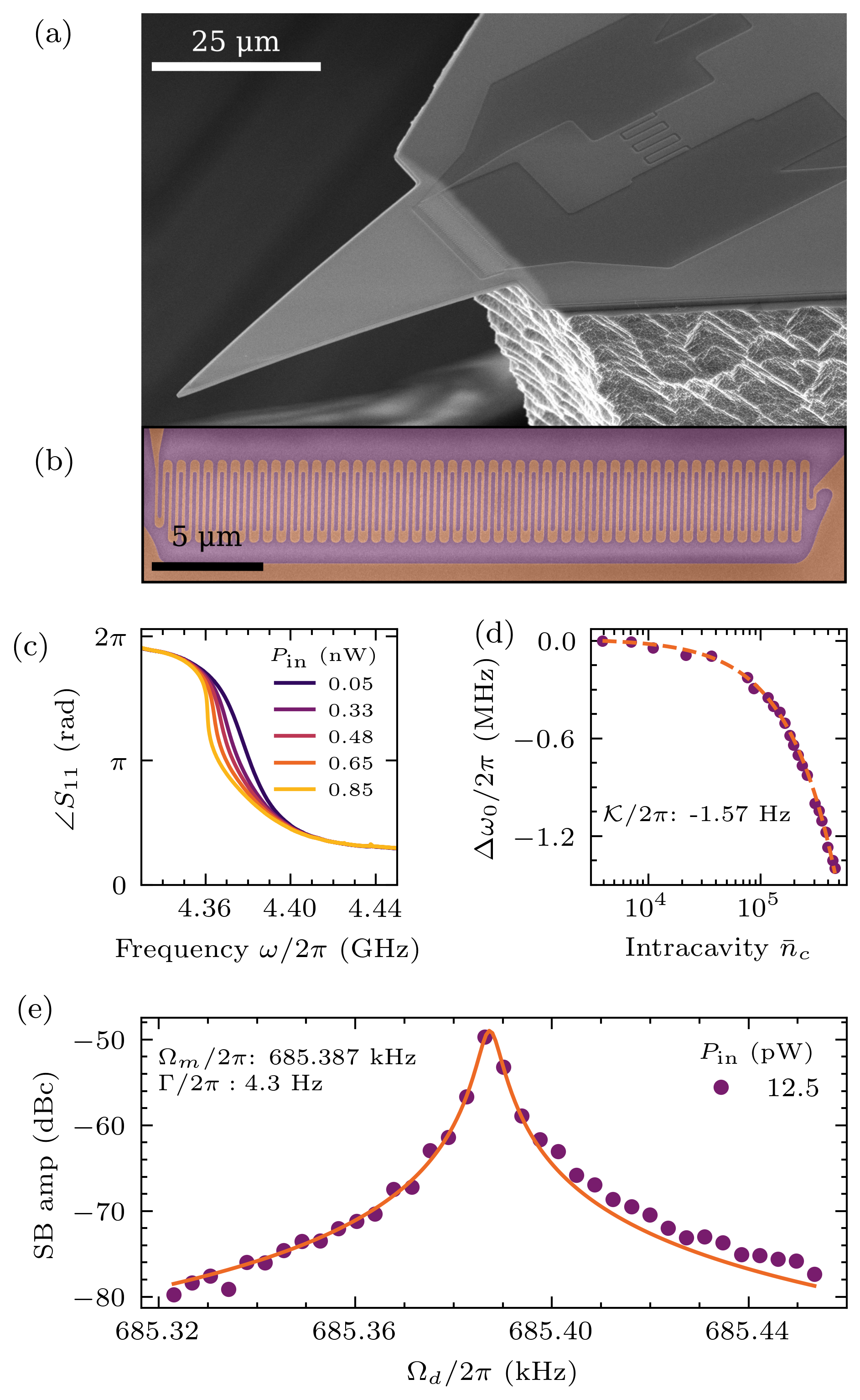}
    \caption{\label{fig:cavity_nonlinearity}
        Scanning electron microscope (SEM) image of (a) the cantilever and (b) the \SI{100}{\nano\metre}-wide meandering nanowire inductor. The cantilever is formed from a \SI{600}{\nano\metre}-thick silicon nitride plate and the nanowire is etched from a \SI{15}{\nano\metre}-thin film of Nb-Ti-N. 
        (c) Phase versus frequency of the microwave resonance for increasing input power $P_{\mathrm{in}}$. The resonance frequency of the cavity $\omega_{0}$ shifts to lower frequencies with increasing $P_{\mathrm{in}}$, typical of a resonator with a Kerr-type nonlinearity.
        (d) The shift in resonance frequency $\Delta\omega_0$ as a function of intracavity photon number $\ncav$ for a pump blue-detuned by approximately \SI{50}{\mega\hertz}. For each pump power, $\omega_0$ is measured by sweeping a weak probe tone through resonance. The Kerr coefficient is $\Kerr / 2 \pi$ = \SI{-1.57}{\hertz\per photon}. (e) Mechanical susceptibility measured from the driven motional sideband with resonant pumping in the linear regime of the cavity.  The sideband amplitude is expressed in decibel with respect to the measured response at the carrier pump frequency (dBc).
    }
\end{figure}

Cantilevers operating in cryogenic environments typically have mechanical resonance frequencies that put the mechanical mode in a thermal state with $\nmech \approx \kB T / \hbar\Omega_{m} \gg 1$. Figure~\ref{fig:standard_quantum_limit} shows that, compared to the standard quantum limit, the thermally noise-limited case has an extended interval of power where force sensitivity is nearly constant and solely determined by the properties of the mechanical resonator. However, the nonlinearity of superconducting microwave circuits commonly limit $\ncav$, which, together with a small coupling rate $g_{0}$, result in a situation where force sensitivity is instead limited by the added noise of a subsequent LNA. As shown in Fig.~\ref{fig:standard_quantum_limit}, one may improve the force sensitivity by adding a quantum-limited amplifier between the resonant circuit, hereafter called ``cavity'', and the LNA. For a sufficiently high-gain phase-preserving quantum-limited parametric amplifier, the number of added photons by the amplifier to the measurement is given by $n_\mathrm{PA} = 1/2$ ~\cite{caves1982}. Yet adding a separate quantum-limited amplifier comes at significant cost in complexity and more complicated operation, including an additional pump, pump-cancellation tone, as well as additional isolators and associated cabling~\cite{teufel2011sidebandcooling}. In our work, we investigate a simpler implementation where a nonlinear cavity is not only used for transduction of motion to measured signal, but also for amplification, i.e. with one pump both generating and amplifying the motional sidebands.

\begin{figure}
    \centering
    \includegraphics[width=\linewidth]{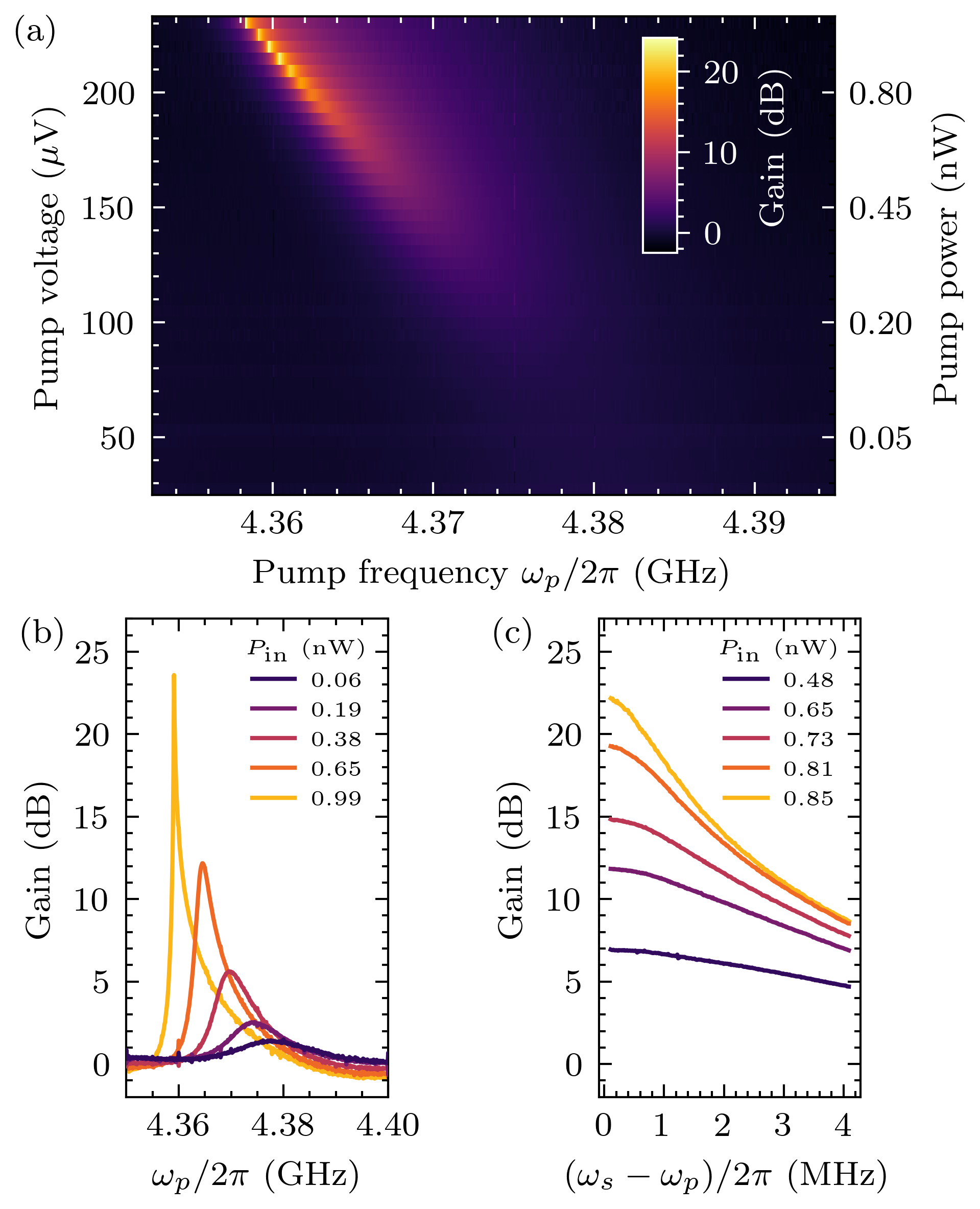}
    \caption{\label{fig:signal_parametric_amplification}
        (a) Four-wave mixing parametric gain of an injected signal tone $\omega_{s}$ as a function of pump frequency $\omega_{p}$ and pump power $P_{\mathrm{in}}$ at fixed detuning $(\omega_{s}- \omega_{p})/2\pi$ = \SI{685.386}{\kilo\hertz}.
        (b) Signal gain versus $\omega_{p}$ at various pump power.
        (c) Gain versus $\omega_{s}$ for various $\omega_{p}$ and $P_{\mathrm{in}}$ along the ridge of maximum gain in (a).
    }
\end{figure}

We demonstrate this implementation with the device shown in Fig.~\ref{fig:cavity_nonlinearity}(a). The cantilever force sensor is tightly integrated with a compact microwave circuit consisting an interdigitated capacitor in series with a long meandering nanowire having large kinetic inductance.  The nanowire has width \SI{100}{\nano\meter} and thickness \SI{15}{\nano\meter}, and it meanders along the base of the cantilever, as shown in detail in the false-color micrograph Fig.~\ref{fig:cavity_nonlinearity}(b). The silicon nitride (Si-N) micro-cantilever has a fundamental bending mode with resonance frequency $\Omega_{m} / 2 \pi = \SI{685.387}{\kilo\hertz}$ and linewidth $\Gamma/2\pi = \SI{4.3}{\hertz}$. The microwave cavity with resonance frequency $\omega_{0} / 2 \pi = \SI{4.378}{\giga\hertz}$ is strongly overcoupled to the transmission line, having total linewidth $\kappa / 2 \pi = \SI{21.18}{\mega\hertz}$, corresponding to a loaded quality factor of $Q = \num{206.7}$.  Cantilever bending generates surface strain which is maximum at the line where the cantilever meets the silicon substrate. This strain changes the nanowire's kinetic inductance, shifting the cavity resonance and thereby realizing electromechanical mode coupling~\cite{roos2023kimec, scarano2024temperature}. For purely geometric coupling where the kinetic inductance per unit length is assumed constant, simulations of the strain~\cite{roos2024phd} result in a single-photon coupling strength $g_0 \approx \SI{50}{\milli\hertz}$, giving an effective cooperativity of $\Ceff \approx 10^{-4}$ at intracavity photon number $\ncav = 10^{6}$.

Figure~\ref{fig:cavity_nonlinearity}(c) shows the phase response of the cavity as a function of frequency and power measured in a dilution refrigerator at $T=\SI{10}{\milli\kelvin}$. The phase of the reflected signal changes by $2 \pi$ when sweeping through resonance, characteristic of an overcoupled resonator measured in reflection. Increasing power shifts the resonance frequency to lower values and the phase response sharpens, as expected from a current-induced pair-breaking nonlinear inductance. Such behavior is approximated to leading order by a Kerr-type nonlinearity, where the Kerr coefficient $\Kerr$ describes the strength of the nonlinearity in terms of a frequency shift per photon. We measure the shift of the cavity resonance frequency as a function of intracavity photon number $\ncav$, by varying the power of a blue-detuned pump while sweeping a much weaker probe tone through resonance. Figure~\ref{fig:cavity_nonlinearity}(d) shows the result of this measurement and the linear fit to determine a Kerr coefficient $\Kerr / 2 \pi$ = \SI{-1.57}{\hertz\per photon} for our device. 

We operate the nonlinear cavity as a four-wave mixing parametric amplifier~\cite{anferov2020nbn} and analyze the gain $\mathcal{G}$ by injecting a single probe tone at $\omega_{s}$, blue-detuned by $\Omega_{m}$ from a strong pump tone at $\omega_{p}$. The amplifier bandwidth is smaller than the cavity linewidth $\kappa$ and it decreases with increasing gain (pump power). We therefore intentionally designed the cavity with a relatively large $\kappa$ (low $Q$) to operate in the unresolved-sideband regime, ensuring high gain at detuning $\Omega_m$. Figures~\ref{fig:signal_parametric_amplification}(a) and (b) show the measured gain versus pump frequency for various pump powers, obtained by sweeping both pump and probe frequency with fixed separation $\omega_{s} - \omega_{p} = \Omega_{m}$. For increasing pump power the gain peak shifts to lower frequency, as expected for a negative Kerr coefficient. For our device, we reach $\mathcal{G} \approx \SI{24}{\decibel}$ for the largest pump power before the cavity bifurcates and the gain degrades. Figure~\ref{fig:signal_parametric_amplification}(c) shows the gain and bandwidth for selected pump powers, measured by fixing $\omega_{p}$ at the previously determined maximum gain for each power, and sweeping the signal tone $\omega_{s}$. 

\begin{figure*}
    \centering
    \includegraphics[width=\linewidth]{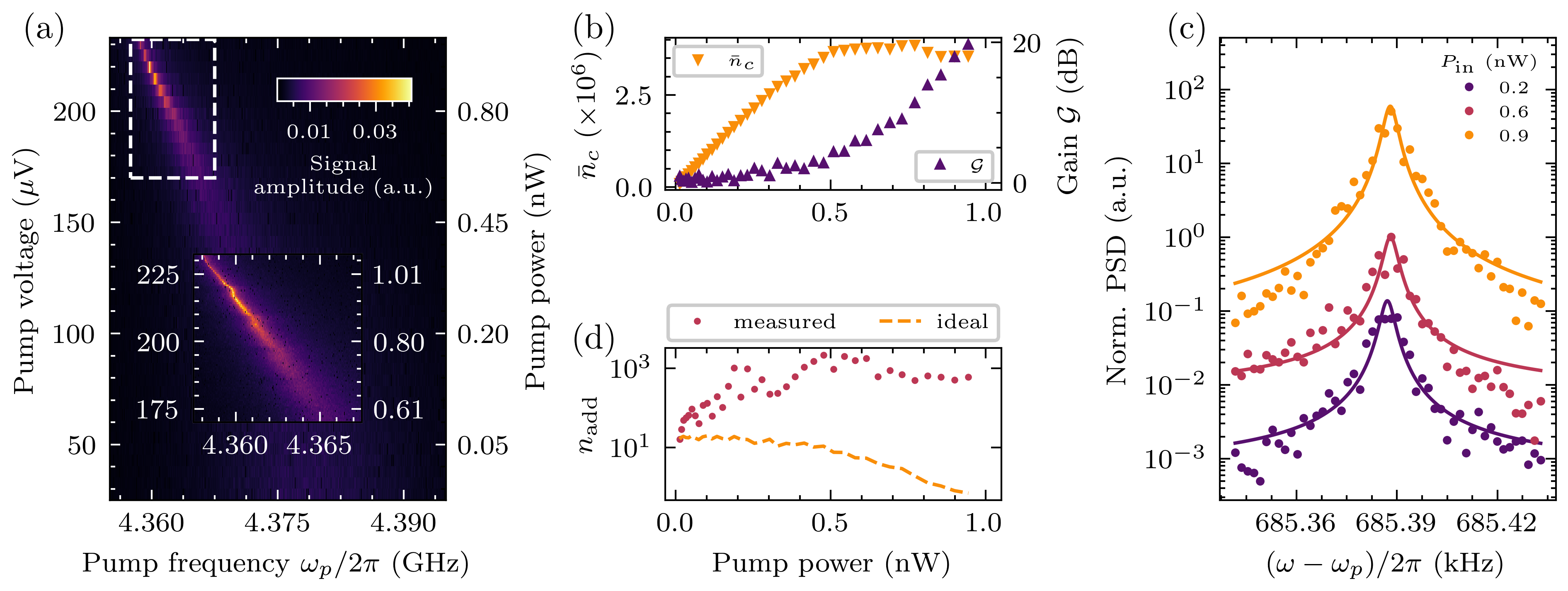}
    \caption{\label{fig:transduction_gain}
        (a) Parametric amplification of an upconverted motional sideband when the mechanical mode is driven with a frequency comb.  The area under the motional sideband is plotted as a function of pump frequency $\omega_{p}$ and pump amplitude/power, normalized to the input amplitude. Increasing the pump power shifts the cavity's resonance frequency and amplifies the response at the motional sideband. (Inset) Zoom of the dashed box with finer stepping, highlighting the ridge of maximum response. 
        (b) Transduction gain $\mathcal{G}$ and $\ncav$ at multiple pump powers along the ridge of maximum sideband response in (a) .
        (c) Fluctuations of the undriven mechanical mode measured at the upper motional sideband for three pump powers with $\omega_{p}$ placed on the ridge of maximum gain. The power spectral density (PSD) is normalized to the intracavity photon number $\ncav$.
        (d) Measured $\nadd$ and that expected for a quantum limited parametric amplifier followed by the LNA.
    }
\end{figure*}

To demonstrate transduction gain, we repeat the sweep over pump power and pump frequency, now driving the cantilever through a piezoelectric shaker. The cantilever motion generates sidebands in the upconverted spectrum, at either side of $\omega_{p}$. Since two sidebands are generated, the gain will be sensitive to their relative phase, which is determined by the properties of the cavity. However, in the unresolved-sideband regime, this relative phase is such that transduction gain is observed~\cite{hatridge2011dispersive}. To capture the motion spectrum we use a multifrequency lockin to excite the cantilever with a tuned frequency comb with equal amplitudes, and phases chosen to reduce peak excitation in the time domain (see Supplementary Material).  The lockin measures the response at the upconverted comb frequencies, capturing the mechanical motion spectrum in a single measurement time window. Fitting a model composed of a Lorentzian with a frequency-independent added noise to the data, we extract the area under the sideband, proportional to the amplitude spectrum of the drive mechanical displacement.
Fig.~\ref{fig:transduction_gain}(a) shows the integrated sideband response normalized to the input amplitude, from which we identify the frequency of maximum response for each pump power.

Maximum transduction gain follows a ridge in pump power and frequency, as expected for parametric gain generated close to bifurcation. Figure~\ref{fig:transduction_gain}(b) shows the transduction gain $\mathcal{G}$ and $\ncav$ versus pump power, where pump frequency is adjusted to follow the ridge of maximum sideband response.  We determine $\ncav$ using methods described in Ref.~\cite{bothner2022kerroptomechanics}. 
At each pump power, the amplified signal is the integrated sideband power spectrum, normalized to $\ncav$. The transduction gain is given by the ratio of this amplified signal, to that in the low power regime, where $\mathcal{G} \equiv 1$.  For a detailed description see the supplementary material.

Operating the sensor on this ridge of maximum gain, we are able to resolve the amplified motional noise of the undriven cantilever, as show in Fig.~\ref{fig:transduction_gain}(c). The plotted data are an average of ten consecutive Power Spectral Densities (PSDs) measured with resolution bandwidth \SI{1.86}{\hertz}. The solid lines represent the best fit of a model comprised of a Lorentzian plus white added noise. Averaging over 100 consecutive fits, we extract the added noise expressed as equivalent photon number, as shown in Fig.~\ref{fig:transduction_gain}(d).

The ideal phase-insensitive parametric amplifier would result in a reduction of the added noise with increasing gain, approaching $\nadd = 1/2$ at high gain [dashed curve in Fig. \ref{fig:transduction_gain}(d)] as described by 
\begin{equation}
    \nadd = n_{\mathrm{PA}}  + \frac{n_{\mathrm{LNA}}}{\mathcal{G}}.
    \label{eqn:total_added_noise}
\end{equation}
However, we observe instead an increase in the added noise. Possible explanations of this additional noise could be nonlinear loss mechanisms in the cavity, for example increased quasiparticle losses associated with the current-induced pair-breaking nonlinearity, or heating of the cavity by the pump~\cite{malnou2021twpa}.

The use of a nonlinear cavity to parametrically amplify the motional sidebands  also results in increased backaction noise, due to the amplification of the intracavity fluctuations~\cite{rodrigues2022kerramplifier}. In contrast, a cold isolator screens the cavity from this backaction when using a separate parametric amplifier~\cite{teufel2009nanomechanicalmotion}. The potential improvements in the resulting force sensitivity are contingent upon the effective cooperativity.  In the case considered here $\Ceff \ll 1/4$ such amplified backaction is negligible at the observed level of gain.

In conclusion, we described a micro-cantilever force-sensor with a compact and integrated microwave cavity. The sensor employed kinetic inductive electromechanical coupling to realize force transduction, and the nonlinearity of the superconducting cavity was used for parametric amplification. The measurement configuration required only a single pump to combine these two effects, achieving up to \SI{21}{\decibel} transduction gain of the electromechanical motional sidebands. However, noise analysis revealed that this gain did not come with an improvement in signal-to-noise ratio, most likely due to nonlinear loss mechanisms or heating by the pump. It is possible to mitigate these effects by designing for higher Kerr coefficient through a thinner superconducting film (lower critical current) or through a shorter nanowire (larger current for given $\ncav$). Further study of this sensor concept is required to determine under which circumstances this gain mechanism can be exploited for improved force sensitivity. 

\section*{Supplementary Material}
See the Supplementary Material for details on the multifrequency lockin measurement technique with a tuned frequency comb driving the mechanical oscillator, determination of the $\Kerr$ coefficient with two tone spectroscopy and analysis of the intracavity photon number and transduction gain.

\section*{Acknowledgements}

We acknowledge funding from the European Union Horizon 2020 EIC Pathfinder Grant Agreement No. 828966 --- QAFM and the Swedish SSF Grant No. ITM17--0343 supported this work. We thank the Quantum-Limited Atomic Force Microscopy (QAFM) team for fruitful discussions: T. Glatzel, M. Zutter, E. Tholén, D. Forchheimer, I. Ignat, M. Kwon, and D. Platz.  We particularly thank anonymous reviewer 2 whose careful reading and thoughtful comments helped to substantially improve the paper.

\section*{Author Declarations}

\subsection*{Conflict of Interest}
The authors have no conflicts to disclose.

\subsection*{Author Contributions}
\textbf{Ermes Scarano:} Conceptualization (equal); Investigation (lead); Formal Analysis (equal); Visualization (equal); Writing -- original draft (lead); Writing -- review and editing (equal).
\textbf{Elisabet K. Arvidsson:} Conceptualization (equal); Investigation (supporting); Formal Analysis (equal); Visualization (equal); Writing -- original draft (supporting); Writing -- review and editing (equal).
\textbf{August K. Roos:} Conceptualization (equal); Investigation (supporting); Formal Analysis (supporting); Visualization (equal); Writing -- original draft (supporting); Writing -- review and editing (supporting).
\textbf{Erik Holmgren:} Supervision (supporting).
\textbf{David B. Haviland:} Conceptualization (equal); Funding acquisition (lead); Resources (lead); Project administration (lead); Supervision (lead);  Writing -- original draft (supporting); Writing -- review and editing (equal).

\section*{Data Availability}
The data that support the findings of this study are openly available in Zenodo at \url{https://doi.org/10.5281/zenodo.11656972}, reference number 11656972.

\providecommand{\noopsort}[1]{}\providecommand{\singleletter}[1]{#1}%


\begin{thebibliography}{0}%
\makeatletter
\providecommand \@ifxundefined [1]{%
 \@ifx{#1\undefined}
}%
\providecommand \@ifnum [1]{%
 \ifnum #1\expandafter \@firstoftwo
 \else \expandafter \@secondoftwo
 \fi
}%
\providecommand \@ifx [1]{%
 \ifx #1\expandafter \@firstoftwo
 \else \expandafter \@secondoftwo
 \fi
}%
\providecommand \natexlab [1]{#1}%
\providecommand \enquote  [1]{``#1''}%
\providecommand \bibnamefont  [1]{#1}%
\providecommand \bibfnamefont [1]{#1}%
\providecommand \citenamefont [1]{#1}%
\providecommand \href@noop [0]{\@secondoftwo}%
\providecommand \href [0]{\begingroup \@sanitize@url \@href}%
\providecommand \@href[1]{\@@startlink{#1}\@@href}%
\providecommand \@@href[1]{\endgroup#1\@@endlink}%
\providecommand \@sanitize@url [0]{\catcode `\\12\catcode `\$12\catcode
  `\&12\catcode `\#12\catcode `\^12\catcode `\_12\catcode `\%12\relax}%
\providecommand \@@startlink[1]{}%
\providecommand \@@endlink[0]{}%
\providecommand \url  [0]{\begingroup\@sanitize@url \@url }%
\providecommand \@url [1]{\endgroup\@href {#1}{\urlprefix }}%
\providecommand \urlprefix  [0]{URL }%
\providecommand \Eprint [0]{\href }%
\providecommand \doibase [0]{https://doi.org/}%
\providecommand \selectlanguage [0]{\@gobble}%
\providecommand \bibinfo  [0]{\@secondoftwo}%
\providecommand \bibfield  [0]{\@secondoftwo}%
\providecommand \translation [1]{[#1]}%
\providecommand \BibitemOpen [0]{}%
\providecommand \bibitemStop [0]{}%
\providecommand \bibitemNoStop [0]{.\EOS\space}%
\providecommand \EOS [0]{\spacefactor3000\relax}%
\providecommand \BibitemShut  [1]{\csname bibitem#1\endcsname}%
\let\auto@bib@innerbib\@empty
\end{thebibliography}%


\providecommand{\noopsort}[1]{}\providecommand{\singleletter}[1]{#1}%
\begin{thebibliography}{2}
\providecommand{\natexlab}[1]{#1}
\providecommand{\url}[1]{\texttt{#1}}
\expandafter\ifx\csname urlstyle\endcsname\relax
  \providecommand{\doi}[1]{doi: #1}\else
  \providecommand{\doi}{doi: \begingroup \urlstyle{rm}\Url}\fi

\bibitem[Newman(1965)]{newman1965}
D.~J. Newman.
\newblock An $l_1$ extremal problem for polynomials.
\newblock \emph{Proceedings of the American Mathematical Society}, 16\penalty0
  (6):\penalty0 1287--1290, 1965.
\newblock ISSN 0002-9939.
\newblock \doi{10.1090/S0002-9939-1971-0280688-0}.

\bibitem[Bothner et~al.(2022)Bothner, Rodrigues, and
  Steele]{bothner2022kerroptomechanics}
Daniel Bothner, Ines~C. Rodrigues, and Gary~A. Steele.
\newblock Four-wave-cooling to the single phonon level in kerr optomechanics.
\newblock \emph{Communications Physics}, 5\penalty0 (1):\penalty0 33, Feb 2022.
\newblock ISSN 2399-3650.
\newblock \doi{10.1038/s42005-022-00808-3}.
\newblock URL \url{https://doi.org/10.1038/s42005-022-00808-3}.

\end{thebibliography}


\begin{thebibliography}{32}
\providecommand{\natexlab}[1]{#1}
\providecommand{\url}[1]{\texttt{#1}}
\expandafter\ifx\csname urlstyle\endcsname\relax
  \providecommand{\doi}[1]{doi: #1}\else
  \providecommand{\doi}{doi: \begingroup \urlstyle{rm}\Url}\fi

\bibitem[Teufel et~al.(2009)Teufel, Donner, Castellanos-Beltran, Harlow, and
  Lehnert]{teufel2009nanomechanicalmotion}
J.~D. Teufel, T.~Donner, M.~A. Castellanos-Beltran, J.~W. Harlow, and K.~W.
  Lehnert.
\newblock Nanomechanical motion measured with an imprecision below that at the
  standard quantum limit.
\newblock \emph{Nature Nanotechnology}, 4\penalty0 (12):\penalty0 820--823, Dec
  2009.
\newblock ISSN 1748-3395.
\newblock \doi{10.1038/nnano.2009.343}.
\newblock URL \url{https://doi.org/10.1038/nnano.2009.343}.

\bibitem[Regal et~al.(2008)Regal, Teufel, and
  Lehnert]{regal2008nanomechanicalmotion}
C.~A. Regal, J.~D. Teufel, and K.~W. Lehnert.
\newblock Measuring nanomechanical motion with a microwave cavity
  interferometer.
\newblock \emph{Nature Physics}, 4\penalty0 (7):\penalty0 555--560, Jul 2008.
\newblock ISSN 1745-2481.
\newblock \doi{10.1038/nphys974}.
\newblock URL \url{https://doi.org/10.1038/nphys974}.

\bibitem[Aspelmeyer et~al.(2014)Aspelmeyer, Kippenberg, and
  Marquardt]{aspelmeyer2014review}
Markus Aspelmeyer, Tobias~J. Kippenberg, and Florian Marquardt.
\newblock Cavity optomechanics.
\newblock \emph{Rev. Mod. Phys.}, 86:\penalty0 1391--1452, Dec 2014.
\newblock \doi{10.1103/RevModPhys.86.1391}.
\newblock URL \url{https://link.aps.org/doi/10.1103/RevModPhys.86.1391}.

\bibitem[Nation et~al.(2016)Nation, Suh, and
  Blencowe]{nation2016ultrastrongcoupling}
P.~D. Nation, J.~Suh, and M.~P. Blencowe.
\newblock Ultrastrong optomechanics incorporating the dynamical casimir effect.
\newblock \emph{Phys. Rev. A}, 93:\penalty0 022510, Feb 2016.
\newblock \doi{10.1103/PhysRevA.93.022510}.
\newblock URL \url{https://link.aps.org/doi/10.1103/PhysRevA.93.022510}.

\bibitem[Reed et~al.(2017)Reed, Mayer, Teufel, Burkhart, Pfaff, Reagor,
  Sletten, Ma, Schoelkopf, Knill, and Lehnert]{reed2017strongcoupling}
A.~P. Reed, K.~H. Mayer, J.~D. Teufel, L.~D. Burkhart, W.~Pfaff, M.~Reagor,
  L.~Sletten, X.~Ma, R.~J. Schoelkopf, E.~Knill, and K.~W. Lehnert.
\newblock Faithful conversion of propagating quantum information to mechanical
  motion.
\newblock \emph{Nature Physics}, 13\penalty0 (12):\penalty0 1163--1167, Dec
  2017.
\newblock ISSN 1745-2481.
\newblock \doi{10.1038/nphys4251}.
\newblock URL \url{https://doi.org/10.1038/nphys4251}.

\bibitem[Rodrigues et~al.(2019)Rodrigues, Bothner, and
  Steele]{rodrigues2019couplingmicrowavephotons}
I.~C. Rodrigues, D.~Bothner, and G.~A. Steele.
\newblock Coupling microwave photons to a mechanical resonator using quantum
  interference.
\newblock \emph{Nature Communications}, 10\penalty0 (1):\penalty0 5359, Nov
  2019.
\newblock ISSN 2041-1723.
\newblock \doi{10.1038/s41467-019-12964-2}.
\newblock URL \url{https://doi.org/10.1038/s41467-019-12964-2}.

\bibitem[Zoepfl et~al.(2020)Zoepfl, Juan, Schneider, and
  Kirchmair]{zoepfl2020singlephotoncooling}
D.~Zoepfl, M.~L. Juan, C.~M.~F. Schneider, and G.~Kirchmair.
\newblock Single-photon cooling in microwave magnetomechanics.
\newblock \emph{Phys. Rev. Lett.}, 125:\penalty0 023601, Jul 2020.
\newblock \doi{10.1103/PhysRevLett.125.023601}.
\newblock URL \url{https://link.aps.org/doi/10.1103/PhysRevLett.125.023601}.

\bibitem[Peterson et~al.(2019)Peterson, Kotler, Lecocq, Cicak, Jin, Simmonds,
  Aumentado, and Teufel]{peterson2019strongcoupling}
G.~A. Peterson, S.~Kotler, F.~Lecocq, K.~Cicak, X.~Y. Jin, R.~W. Simmonds,
  J.~Aumentado, and J.~D. Teufel.
\newblock Ultrastrong parametric coupling between a superconducting cavity and
  a mechanical resonator.
\newblock \emph{Phys. Rev. Lett.}, 123:\penalty0 247701, Dec 2019.
\newblock \doi{10.1103/PhysRevLett.123.247701}.
\newblock URL \url{https://link.aps.org/doi/10.1103/PhysRevLett.123.247701}.

\bibitem[Ren et~al.(2020)Ren, Matheny, MacCabe, Luo, Pfeifer, Mirhosseini, and
  Painter]{ren2020highquantumcooperativity}
Hengjiang Ren, Matthew~H. Matheny, Gregory~S. MacCabe, Jie Luo, Hannes Pfeifer,
  Mohammad Mirhosseini, and Oskar Painter.
\newblock Two-dimensional optomechanical crystal cavity with high quantum
  cooperativity.
\newblock \emph{Nature Communications}, 11\penalty0 (1):\penalty0 3373, Jul
  2020.
\newblock ISSN 2041-1723.
\newblock \doi{10.1038/s41467-020-17182-9}.
\newblock URL \url{https://doi.org/10.1038/s41467-020-17182-9}.

\bibitem[Bozkurt et~al.(2023)Bozkurt, Zhao, Joshi, LeDuc, Day, and
  Mirhosseini]{bozkurt2023quantum}
Alkim Bozkurt, Han Zhao, Chaitali Joshi, Henry~G. LeDuc, Peter~K. Day, and
  Mohammad Mirhosseini.
\newblock A quantum electromechanical interface for long-lived phonons.
\newblock \emph{Nature Physics}, 19\penalty0 (9):\penalty0 1326--1332, Sep
  2023.
\newblock ISSN 1745-2481.
\newblock \doi{10.1038/s41567-023-02080-w}.
\newblock URL \url{https://doi.org/10.1038/s41567-023-02080-w}.

\bibitem[Cattiaux et~al.(2020)Cattiaux, Zhou, Kumar, Golokolenov, Gazizulin,
  Luck, de~L\'epinay, Sillanp\"a\"a, Armour, Fefferman, and
  Collin]{cattiaux2020beyondlinearcoupling}
D.~Cattiaux, X.~Zhou, S.~Kumar, I.~Golokolenov, R.~R. Gazizulin, A.~Luck,
  L.~Mercier de~L\'epinay, M.~Sillanp\"a\"a, A.~D. Armour, A.~Fefferman, and
  E.~Collin.
\newblock Beyond linear coupling in microwave optomechanics.
\newblock \emph{Phys. Rev. Res.}, 2:\penalty0 033480, Sep 2020.
\newblock \doi{10.1103/PhysRevResearch.2.033480}.
\newblock URL \url{https://link.aps.org/doi/10.1103/PhysRevResearch.2.033480}.

\bibitem[Shin et~al.(2022)Shin, Ryu, Miri, Shim, Choi, Alù, Suh, and
  Cha]{shin2022frequencycombs}
Junghyun Shin, Younghun Ryu, Mohammad-Ali Miri, Seung-Bo Shim, Hyoungsoon Choi,
  Andrea Alù, Junho Suh, and Jinwoong Cha.
\newblock On-chip microwave frequency combs in a superconducting
  nanoelectromechanical device.
\newblock \emph{Nano Letters}, 22\penalty0 (13):\penalty0 5459--5465, 2022.
\newblock \doi{10.1021/acs.nanolett.2c01503}.
\newblock URL \url{https://doi.org/10.1021/acs.nanolett.2c01503}.
\newblock PMID: 35708318.

\bibitem[Blencowe and Buks(2007)]{blencowe2007squidanalysis}
M.~P. Blencowe and E.~Buks.
\newblock Quantum analysis of a linear dc squid mechanical displacement
  detector.
\newblock \emph{Phys. Rev. B}, 76:\penalty0 014511, Jul 2007.
\newblock \doi{10.1103/PhysRevB.76.014511}.
\newblock URL \url{https://link.aps.org/doi/10.1103/PhysRevB.76.014511}.

\bibitem[Teufel et~al.(2011)Teufel, Donner, Li, Harlow, Allman, Cicak, Sirois,
  Whittaker, Lehnert, and Simmonds]{teufel2011sidebandcooling}
J.~D. Teufel, T.~Donner, Dale Li, J.~W. Harlow, M.~S. Allman, K.~Cicak, A.~J.
  Sirois, J.~D. Whittaker, K.~W. Lehnert, and R.~W. Simmonds.
\newblock Sideband cooling of micromechanical motion to the quantum ground
  state.
\newblock \emph{Nature}, 475\penalty0 (7356):\penalty0 359--363, Jul 2011.
\newblock ISSN 1476-4687.
\newblock \doi{10.1038/nature10261}.
\newblock URL \url{https://doi.org/10.1038/nature10261}.

\bibitem[Safavi-Naeini et~al.(2012)Safavi-Naeini, Chan, Hill, Alegre, Krause,
  and Painter]{safavi-naeini2012nanomechanicalresonator}
Amir~H. Safavi-Naeini, Jasper Chan, Jeff~T. Hill, Thiago P.~Mayer Alegre, Alex
  Krause, and Oskar Painter.
\newblock Observation of quantum motion of a nanomechanical resonator.
\newblock \emph{Phys. Rev. Lett.}, 108:\penalty0 033602, Jan 2012.
\newblock \doi{10.1103/PhysRevLett.108.033602}.
\newblock URL \url{https://link.aps.org/doi/10.1103/PhysRevLett.108.033602}.

\bibitem[Yuan et~al.(2015)Yuan, Singh, Blanter, and Steele]{yuan2015cavity3d}
Mingyun Yuan, Vibhor Singh, Yaroslav~M. Blanter, and Gary~A. Steele.
\newblock Large cooperativity and microkelvin cooling with a three-dimensional
  optomechanical cavity.
\newblock \emph{Nature Communications}, 6\penalty0 (1):\penalty0 8491, 10 2015.
\newblock ISSN 2041-1723.
\newblock \doi{10.1038/ncomms9491}.
\newblock URL \url{https://doi.org/10.1038/ncomms9491}.

\bibitem[Cattiaux et~al.(2021)Cattiaux, Golokolenov, Kumar, Sillanp{\"a}{\"a},
  Mercier~de L{\'e}pinay, Gazizulin, Zhou, Armour, Bourgeois, Fefferman, and
  Collin]{cattiaux2021groundstatecooling}
D.~Cattiaux, I.~Golokolenov, S.~Kumar, M.~Sillanp{\"a}{\"a}, L.~Mercier~de
  L{\'e}pinay, R.~R. Gazizulin, X.~Zhou, A.~D. Armour, O.~Bourgeois,
  A.~Fefferman, and E.~Collin.
\newblock A macroscopic object passively cooled into its quantum ground state
  of motion beyond single-mode cooling.
\newblock \emph{Nature Communications}, 12\penalty0 (1):\penalty0 6182, Oct
  2021.
\newblock ISSN 2041-1723.
\newblock \doi{10.1038/s41467-021-26457-8}.
\newblock URL \url{https://doi.org/10.1038/s41467-021-26457-8}.

\bibitem[Bothner et~al.(2022)Bothner, Rodrigues, and
  Steele]{bothner2022kerroptomechanics}
Daniel Bothner, Ines~C. Rodrigues, and Gary~A. Steele.
\newblock Four-wave-cooling to the single phonon level in kerr optomechanics.
\newblock \emph{Communications Physics}, 5\penalty0 (1):\penalty0 33, Feb 2022.
\newblock ISSN 2399-3650.
\newblock \doi{10.1038/s42005-022-00808-3}.
\newblock URL \url{https://doi.org/10.1038/s42005-022-00808-3}.

\bibitem[Seis et~al.(2022)Seis, Capelle, Langman, Saarinen, Planz, and
  Schliesser]{seis2022cooling}
Yannick Seis, Thibault Capelle, Eric Langman, Sampo Saarinen, Eric Planz, and
  Albert Schliesser.
\newblock Ground state cooling of an ultracoherent electromechanical system.
\newblock \emph{Nature Communications}, 13\penalty0 (1):\penalty0 1507, Mar
  2022.
\newblock ISSN 2041-1723.
\newblock \doi{10.1038/s41467-022-29115-9}.
\newblock URL \url{https://doi.org/10.1038/s41467-022-29115-9}.

\bibitem[Zoepfl et~al.(2023)Zoepfl, Juan, Diaz-Naufal, Schneider, Deeg,
  Sharafiev, Metelmann, and Kirchmair]{zoepfl2023kerrenhancedcooling}
D.~Zoepfl, M.~L. Juan, N.~Diaz-Naufal, C.~M.~F. Schneider, L.~F. Deeg,
  A.~Sharafiev, A.~Metelmann, and G.~Kirchmair.
\newblock Kerr enhanced backaction cooling in magnetomechanics.
\newblock \emph{Phys. Rev. Lett.}, 130:\penalty0 033601, Jan 2023.
\newblock \doi{10.1103/PhysRevLett.130.033601}.
\newblock URL \url{https://link.aps.org/doi/10.1103/PhysRevLett.130.033601}.

\bibitem[Zmuidzinas(2012)]{zmuidzinas2012review}
Jonas Zmuidzinas.
\newblock Superconducting microresonators: Physics and applications.
\newblock \emph{Annual Review of Condensed Matter Physics}, 3\penalty0
  (1):\penalty0 169--214, 2012.
\newblock \doi{10.1146/annurev-conmatphys-020911-125022}.
\newblock URL \url{https://doi.org/10.1146/annurev-conmatphys-020911-125022}.

\bibitem[Vissers et~al.(2016)Vissers, Erickson, Ku, Vale, Wu, Hilton, and
  Pappas]{vissers2016twpa}
M.~R. Vissers, R.~P. Erickson, H.-S. Ku, Leila Vale, Xian Wu, G.~C. Hilton, and
  D.~P. Pappas.
\newblock {Low-noise kinetic inductance traveling-wave amplifier using
  three-wave mixing}.
\newblock \emph{Applied Physics Letters}, 108\penalty0 (1):\penalty0 012601, 01
  2016.
\newblock ISSN 0003-6951.
\newblock \doi{10.1063/1.4937922}.
\newblock URL \url{https://doi.org/10.1063/1.4937922}.

\bibitem[Erickson and Pappas(2017)]{erickson2017theorykinetictwpa}
R.~P. Erickson and D.~P. Pappas.
\newblock Theory of multiwave mixing within the superconducting
  kinetic-inductance traveling-wave amplifier.
\newblock \emph{Phys. Rev. B}, 95:\penalty0 104506, Mar 2017.
\newblock \doi{10.1103/PhysRevB.95.104506}.
\newblock URL \url{https://link.aps.org/doi/10.1103/PhysRevB.95.104506}.

\bibitem[Bowen and Milburn(2015)]{bowen2015quantumoptomechanics}
W.P. Bowen and G.J. Milburn.
\newblock \emph{Quantum Optomechanics}.
\newblock Taylor \& Francis, Boca Raton, first edition, 2015.
\newblock ISBN 9781482259155.
\newblock URL \url{https://doi.org/10.1201/b19379}.

\bibitem[Caves(1982)]{caves1982}
Carlton~M. Caves.
\newblock Quantum limits on noise in linear amplifiers.
\newblock \emph{Phys. Rev. D}, 26:\penalty0 1817--1839, Oct 1982.
\newblock \doi{10.1103/PhysRevD.26.1817}.
\newblock URL \url{https://link.aps.org/doi/10.1103/PhysRevD.26.1817}.

\bibitem[Roos et~al.(2023)Roos, Scarano, Arvidsson, Holmgren, and
  Haviland]{roos2023kimec}
August~K. Roos, Ermes Scarano, Elisabet~K. Arvidsson, Erik Holmgren, and
  David~B. Haviland.
\newblock Kinetic inductive electromechanical transduction for nanoscale force
  sensing.
\newblock \emph{Phys. Rev. Appl.}, 20:\penalty0 024022, 08 2023.
\newblock \doi{10.1103/PhysRevApplied.20.024022}.
\newblock URL \url{https://link.aps.org/doi/10.1103/PhysRevApplied.20.024022}.

\bibitem[Scarano et~al.(2024)Scarano, Arvidsson, Roos, Holmgren, and
  Haviland]{scarano2024temperature}
Ermes Scarano, Elisabet~K. Arvidsson, August~K. Roos, Erik Holmgren, and
  David~B. Haviland.
\newblock Temperature dependence of microwave losses in lumped-element
  resonators made from superconducting nanowires with high kinetic inductance,
  2024.

\bibitem[Roos(2024)]{roos2024phd}
August~K. Roos.
\newblock \emph{Superconducting kinetic inductance devices for nanoscale force
  sensing}.
\newblock PhD thesis, KTH, Nanostructure Physics, 2024.
\newblock QC 2024-02-06.

\bibitem[Anferov et~al.(2020)Anferov, Suleymanzade, Oriani, Simon, and
  Schuster]{anferov2020nbn}
Alexander Anferov, Aziza Suleymanzade, Andrew Oriani, Jonathan Simon, and
  David~I. Schuster.
\newblock Millimeter-wave four-wave mixing via kinetic inductance for quantum
  devices.
\newblock \emph{Phys. Rev. Appl.}, 13:\penalty0 024056, Feb 2020.
\newblock \doi{10.1103/PhysRevApplied.13.024056}.
\newblock URL \url{https://link.aps.org/doi/10.1103/PhysRevApplied.13.024056}.

\bibitem[Hatridge et~al.(2011)Hatridge, Vijay, Slichter, Clarke, and
  Siddiqi]{hatridge2011dispersive}
M.~Hatridge, R.~Vijay, D.~H. Slichter, John Clarke, and I.~Siddiqi.
\newblock Dispersive magnetometry with a quantum limited squid parametric
  amplifier.
\newblock \emph{Phys. Rev. B}, 83:\penalty0 134501, Apr 2011.
\newblock \doi{10.1103/PhysRevB.83.134501}.
\newblock URL \url{https://link.aps.org/doi/10.1103/PhysRevB.83.134501}.

\bibitem[Malnou et~al.(2021)Malnou, Vissers, Wheeler, Aumentado, Hubmayr,
  Ullom, and Gao]{malnou2021twpa}
M.~Malnou, M.R. Vissers, J.D. Wheeler, J.~Aumentado, J.~Hubmayr, J.N. Ullom,
  and J.~Gao.
\newblock Three-wave mixing kinetic inductance traveling-wave amplifier with
  near-quantum-limited noise performance.
\newblock \emph{PRX Quantum}, 2:\penalty0 010302, Jan 2021.
\newblock \doi{10.1103/PRXQuantum.2.010302}.
\newblock URL \url{https://link.aps.org/doi/10.1103/PRXQuantum.2.010302}.

\bibitem[Rodrigues et~al.(2022)Rodrigues, Steele, and
  Bothner]{rodrigues2022kerramplifier}
Ines~Corveira Rodrigues, Gary~Alexander Steele, and Daniel Bothner.
\newblock Parametrically enhanced interactions and nonreciprocal bath dynamics
  in a photon-pressure kerr amplifier.
\newblock \emph{Science Advances}, 8\penalty0 (34):\penalty0 eabq1690, 2022.
\newblock \doi{10.1126/sciadv.abq1690}.
\newblock URL \url{https://www.science.org/doi/abs/10.1126/sciadv.abq1690}.

\end{thebibliography}
\end{document}


\title{Supplementary Material:\\Intrinsic Kerr amplification for microwave electromechanics}

\author{Ermes Scarano}
\affiliation{Department of Applied Physics, KTH Royal Institute of Technology, Hannes Alfvéns väg 12, SE-114 19 Stockholm, Sweden}
\author{Elisabet K. Arvidsson}
\affiliation{Department of Applied Physics, KTH Royal Institute of Technology, Hannes Alfvéns väg 12, SE-114 19 Stockholm, Sweden}
\author{August K. Roos}
\affiliation{Department of Applied Physics, KTH Royal Institute of Technology, Hannes Alfvéns väg 12, SE-114 19 Stockholm, Sweden}
\author{Erik Holmgren}
\affiliation{Department of Applied Physics, KTH Royal Institute of Technology, Hannes Alfvéns väg 12, SE-114 19 Stockholm, Sweden}
\author{David B. Haviland}
\email{haviland@kth.se}
\affiliation{Department of Applied Physics, KTH Royal Institute of Technology, Hannes Alfvéns väg 12, SE-114 19 Stockholm, Sweden}
\date{June 14, 2024}%

\maketitle

\onecolumngrid

\section*{Supplementary Material 1: Multifrequency lockin measurement}
The linear response function of a system, e.g. the mechanical susceptibility of an eigenmode of a cantilever, can be determined by stepping a single drive tone through resonance while measuring the amplitude and phase of the response at the drive frequency using the lockin technique. To properly resolve the Lorentzian lineshape the step size $\Delta f$ should be smaller or on the order of the linewidth $\gamma_m$, and the signal must be integrated with a bandwidth $\delta f\leq \Delta f$. Tuning the discrete set of stepped frequencies $f_i = n_i \Delta f$, to be integer multiples of the integration bandwidth $\Delta f=m\delta f$, where $n_i,m$ are positive integers, ensures that no Fourier leakage will occur. The measurement time for a set of $N$ frequencies will be $T=N / \delta f$. 

A much faster measurement is achieved by driving the system with a frequency comb, i.e. simultaneously applying all the drive tones at the multiple frequencies $f_i$.  The multifrequency method reduces the measurement time to $T=1/\delta f$. By tuning the frequencies in the comb to be integer multiples of $\delta f$, the drive signal in the time domain is periodic with $T = 1 / \delta f$ and a common reference for the phases of all tones can be defined, as in a Discrete Fourier Transform. 
The response to this multifrequency drive will be identical to stepping a single-frequency drive, as long as the system remains linear under the applied drive signal. 

A frequency comb with large number of tones can drive the system in to a nonlinear regime at that point in time when the waveform peaks, when the amplitudes at all frequencies align with the same phase. The peak value can be reduced with proper choice of the phases. The worst case is that of equal phases for all the tones, where the drive signal in the time domain is a $\sinc$ pulse. For arbitrary number of frequencies, there is no analytical solution to obtain the minimum \emph{crest factor}, defined as the ratio of the peak to the root-mean-square value of the waveform~\cite{newman1965}. To avoid nonlinear response of the cantilever in our measurement, we drove the mechanical resonator with a comb of nine frequencies and we chose the drive phases by generating multiple sets of random phases, selecting the waveform with the lowest crest factor.

\section*{Supplementary Material 2: Determining the Kerr coefficient}
\begin{figure}
    \centering
    \includegraphics[width=\linewidth]{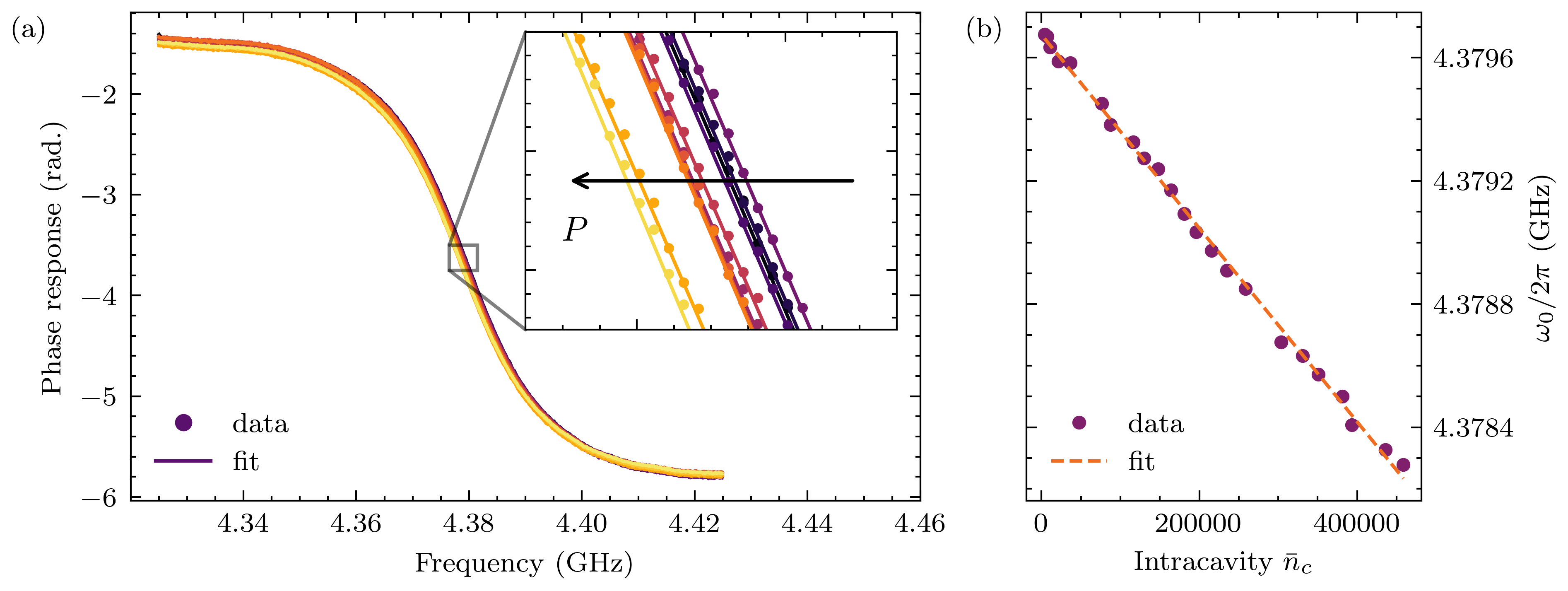}
    \caption{
        (a) Measured response phase versus frequency of the weak probe tone, plotted for various power of a strong blue-detuned pump tone. The pump is fixed at a frequency detuned by \SI{50} {\mega\hertz} from $\omega_0$.  The input pump power $P_\mathrm{in}$  ranges from \SI{-81}{dBm} to \SI{-60}{dBm}.  The inset shows a zoom detailing the small decrease in resonance frequency $\omega_{0}$ for increasing pump power (direction indicated by arrow).
        (b) The shifted resonance frequency as a function of intracavity photons $\ncav$, from which we extract $\Kerr = \SI{-1.572 \pm 0.017}{\hertz\per photon}$
    }
    \label{fig:kerr_fit}
\end{figure}

In Fig.~2(d) of the main text we determine the Kerr coefficient by considering the shift of the cavity's resonance frequency $\omega_{0}$ as a function of intracavity photon number $\ncav$.  
Intracavity photons are produced by a pump tone at the frequency $\omega_p$ which is blue-detuned from $\omega_{0}$  by $\Delta=\SI{50}{\MHz}$, larger than the cavity linewidth $\Delta > \kappa$. 
The corresponding number of intracavity photons circulating in the resonator is given by:
\begin{equation}
    \ncav=\frac{P_{in}}{\hbar\omega_p}\frac{\kex}{\kappa^2/4+\Delta^2}
\end{equation}
The response phase in Fig.~\ref{fig:kerr_fit}(a) is measured by two-tone spectroscopy, with a pump tone at a fixed frequency and a weak probe tone stepping through the cavity resonance. 
The zoom details the shift of resonance frequency with increasing pump power.  The shifted resonance frequency is a linear function of intracavity photon number, with  slope correspondsing to twice the Kerr coefficient $\Kerr$, as shown in Fig.~\ref{fig:kerr_fit}(b).

\section*{Supplementary Material 3: Calculating intracavity photons}
We follow the methods outlined in the Supplementary Methods of Ref.~\cite{bothner2022kerroptomechanics} to determine the number of circulating photons $\ncav$ in the cavity with a Kerr-type nonlinearity. The Kerr effect captures the strength of a nonlinearity in terms the Kerr coefficient $\Kerr$ which we can interpret as the frequency shift per intracavity photon. 
The equation of motion for the intracavity field $a$ for a single port cavity measured in reflection, is given by
\begin{equation}
    \dot{a} = \left[i \left( \omega_{0} + \Kerr \lvert a \rvert^{2} \right) - \frac{\kappa}{2} \right] a + \sqrt{\kex} a_{\mathrm{in}},
    \label{eqn:kerr_eom}
\end{equation}
where $\omega_{0}$ is the bare resonance frequency of the cavity, $\kappa$ its total linewidth, $\kex$ its external linewidth, and $a_{\mathrm{in}}$ photon field at the input of the cavity.

For a single-tone probe at frequency $\omega_{p}$, the input field is given by $a_{\mathrm{in}} = \lvert \alpha_{\mathrm{in}} \rvert \exp(i \omega_{p} t)$. 
In this case, the intracavity field may be written as $a= \alpha \exp(i \omega_{p} t)$. 
Defining a detuning between drive tone and the bare resonance $\Delta = \omega_{p} - \omega_{0}$, the equation of motion may be written,
\begin{equation}
    \alpha \left[ \frac{\kappa}{2} + i \left( \Delta - \Kerr \alpha^{2} \right) \right] = \sqrt{\kex} \lvert \alpha_{\mathrm{in}} \rvert .
    \label{eqn:modified_kerr_eom}
\end{equation}
Multiplying Eqn.~\eqref{eqn:modified_kerr_eom} by its complex conjugate forms  a third-order polynomial in $\ncav = \alpha^*  \alpha$
\begin{equation}
    \Kerr^{2} \ncav^{3} - 2 \Kerr \Delta \ncav^{2} + \left( \Delta^{2} + \frac{\kappa^{2}}{4} \right) \ncav - \kex  \lvert \alpha_{\mathrm{in}} \rvert^{2} = 0,
    \label{eqn:cubic_equation}
\end{equation}
where the number of incoming photons is given by
\begin{equation}
     n_{\mathrm{in}} \equiv \lvert \alpha_{\mathrm{in}} \rvert^{2} = \frac{P_{\mathrm{in}}}{\hbar \omega_{p}} %
     \label{eqn:input_photons}
\end{equation}
For each given value of pump power $P_{\mathrm{in}}$ and drive frequency $\omega_{p}$, and the values of $\Kerr$, and $\omega_{0}$, $\kappa$, $\kex$  previously determined (in the linear regime), we numerically solve the polynomial Eqn.~\eqref{eqn:cubic_equation} to arrive at $\ncav$. 
The polynomial has in general three roots, but our input power corresponds to a regime below bifurcation, where only one real root exists.
With the derived value of the Kerr parameter we can reproduce the $S_{11}(\omega)$ curve for a swept pump tone (Fig.~\ref{fig:nc_v_P_v_f}(a)). For the input-output relation $\langle\alpha_{out}\rangle=\langle\alpha_{in}\rangle-\sqrt{\kex}\langle a\rangle$, the reflection coefficient of the single port cavity is given by
\begin{equation}
    S_{11}=\frac{\langle a_{out}\rangle}{\langle a_{in}\rangle}=1-\frac{\kex}{\kappa/2+i(\Delta-\Kerr a^2)}\label{S11_NL}
\end{equation}
with $\omega_0 / 2\pi = \SI{4.3796}{\GHz}$, $\kappa / 2 \pi \simeq \kex / 2 \pi = \SI{24.186}{\MHz}$, $\Kerr / 2 \pi = \SI{-1.57}{\hertz\per photon}$.

\section*{Supplementary Material 4: Determining the transduction gain}
\begin{figure}
    \centering
    \includegraphics[width=\textwidth]{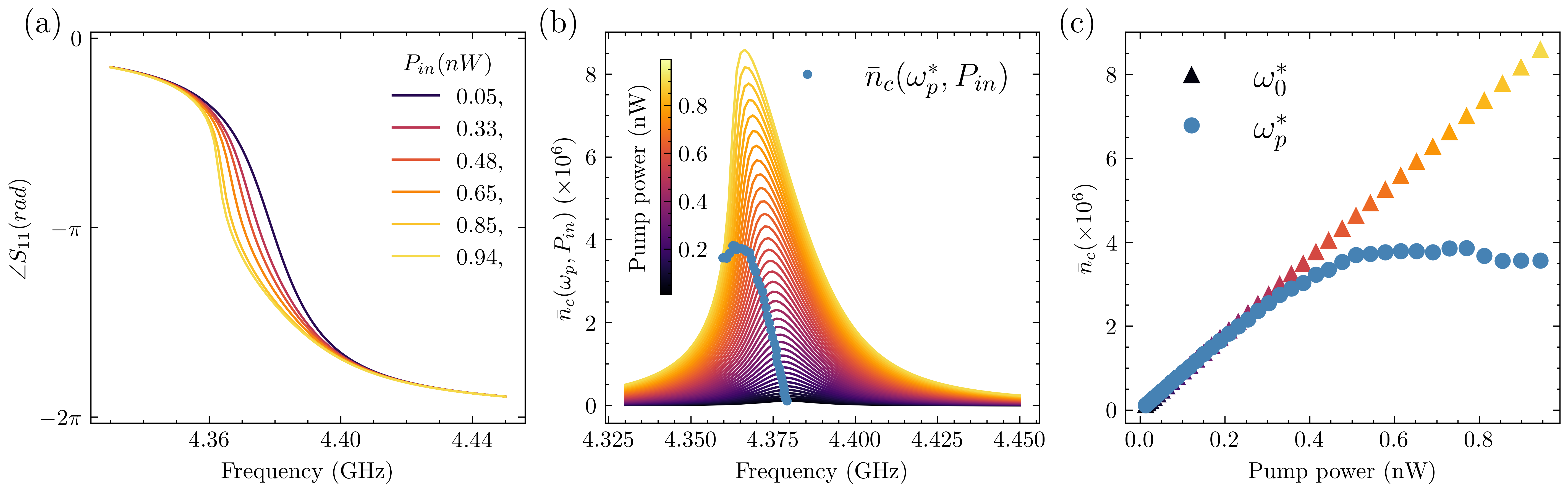}
    \caption{
        (a) Reflection coefficient $S_{11}$ vs frequency simulated from the Kerr model for the single tone solution in Eqn.~\eqref{S11_NL}. This simulations matches well the measurements reported in Fig.~2(c) of the main text.
        (b) The number of intracavity photons $\ncav$  as a function of pump frequency for various input pump power $P_\mathrm{in}$.  The smooth curves are calculated from the Kerr model.  For each pump power a blue point marks the frequency $\omega_p^*$   corresponding to the measured maximum sideband signal.
        (c) Comparison between maximum $\ncav$ and $\ncav$ for a pump frequency that corresponds to maximum gain.
    }
    \label{fig:nc_v_P_v_f}
\end{figure}

The transduction gain as defined in the main text corresponds to the net amplification of the motional sideband, or increase in the amplitude of the response at the sideband frequency, in comparison with that which would be observed with a linear cavity. 
To determine the transduction gain from the experimental data (Fig.~4(a) in the main text), we first find the pump frequency $\omega_{p}^{*}$ at which we observe maximum sideband response, for each input pump power.  
This experimentally determined $\omega_{p}^{*}(P_{in})$ can be estimated analytically from the Kerr model as the frequency $\omega_{0}^{*}(P)$ at which the number of intracavity photons is maximum (see Fig.~\ref{fig:nc_v_P_v_f}). 
The experimental $\omega_{p}^{*}(P_{in})$ does not coincide exactly with the analytical $\omega_{0}^{*}(P_{in})$ but is slightly red detuned, as shown in Fig.~\ref{fig:nc_v_P_v_f}(b,c)

To calculate the value of the transduction gain $G$ we assume that the power spectral density $S_{xx}(\Omega)$ of the mechanical displacement (either driven coherently, or with broad band noise) is independent on the microwave pump power and frequency. 
At the frequency of maximum sideband response corresponding to the $j$th input pump power  $\omega_p^*(P_{in,j})$, the voltage power spectral density of the measured motional sidebands scales as:
\begin{equation}
    S_{VV,j} \propto g_0 S_{xx} G_j \bar{n}_c(\omega_p^*,P_{in,j}).
\end{equation}
In the low power regime ($P_{in,j}<\SI{0.2}{\nW}$) both $S_{VV}$ and $\bar{n}_c$ scale linearly with the input power, defining the regime where gain is unity. Therefore,  at the $j$th power the gain is: 
\begin{equation}
   G_j = \frac {S_{VV,j}/\bar{n}_c(\omega_p^*,P_{in,j})}{S_{VV,0}/\bar{n}_c(\omega_p^*,P_{in,0})}.
\end{equation}
where the index $j=0$ represents the low power regime where the $G=1$.

\providecommand{\noopsort}[1]{}\providecommand{\singleletter}[1]{#1}%